# Transition of the thermal boundary layer and plume over an isothermal section-triangular roof: An experimental study

Haoyu Zhai,[1,2,3] Juan F. Torres,[1,†] Yongling Zhao,[4] and Feng Xu[2]

*[1] ANU HEAT Lab, School of Engineering, The Australian National University, Canberra ACT, Australia*

*[2] School of Physical Science and Engineering, Beijing Jiaotong University, Beijing 100044, China*

*[3] Department of Civil and Environmental Engineering, The Hong Kong Polytechnic University, Hong Kong, China*

*[4] Department of Mechanical and Process Engineering, ETH Zürich, Zürich 8093, Switzerland*

The development of thermal boundary layers and plume near a section-triangular roof under different isothermal heating conditions have been the focus of numerous numerical studies. However, flow transition in this type of flow has never been observed experimentally. Here, phase-shifting interferometry and thermistor measurements are employed to experimentally observe and quantify the flow transitions in a buoyancy-driven flow over an isothermal section-triangular roof. Visualisation of temperature contours is conducted across a wide range of Rayleigh numbers from laminar at $10^3$ to chaotic state at $4 \times 10^6$. Power spectral density of the temperature measurements reveals the type of bifurcations developing as the Rayleigh number is increased. This flow transition is characterised as a complex bifurcation route with the presence of two fundamental frequencies, a low and a high frequency. We found that the thermal stratification in the environment plays a significant role in the flow transition. The spatial development of flow is also quantitatively and qualitatively described. In addition to clarifying flow transition in experiments, the work demonstrates the implementation of phase-shifting interferometry and punctual temperature measurements for characterisation of near-field flow over heated surface.

## 1. Introduction

Understanding thermal flows over a heated surface has been the focus of many studies for over a century (Fan et al. 2021). Inclined surface structures such as section-triangular roofs are ubiquitous, hence they have attracted much attention in research due to its fundamental and engineering significance. Under different governing parameters, the thermal flow over the surface appears to be laminar, transitional, or turbulent. In recent decades, this flow has raised significant interest as the development from laminar to turbulent flow regimes has been predicted to have a complex transition route with various types of bifurcations.

Boundary layer theory is a classic method to investigate viscous flow properties adjacent to a heated surface (Prandtl 1942; Patterson & Imberger 1980). Further, studies (Zhai et al. 2018, 2021) show that thermal flows over a section-triangular roof may be described by the thermal boundary layer adjacent to the surface and the plume above the surface. That is, the thermal boundary layer forms over the roof surface, and the fluid in the thermal boundary layer flows downstream driven by buoyancy and separates from the surface at the separation point, thereby forming a downstream plume. The flow is mainly controlled by the Rayleigh number (Ra), Prandtl number (Pr) and inclination angle ($\theta$). Ra is defined as

$$\mathrm{Ra} = \frac{g\beta\Delta T l^3}{\nu\kappa},\tag{1}$$

where $g$ is the acceleration due to gravity, $\beta$ is the thermal expansion coefficient, $\Delta T$ is the temperature difference between the heated surface and environment, $l$ is the characteristic length of heated surface, $\nu$ is the kinematic viscosity, and $\kappa$ is the thermal diffusivity. An increasing Ra favours the transition from laminar to transitional flow (such as periodic flow) and eventually to turbulent flow. The Prandtl number (Pr) is defined as

$$\mathrm{Pr} = \frac{\nu}{\kappa}.\tag{2}$$

---

† Corresponding author: felipe.torres@anu.edu.au

In addition to the controlling parameters, other parameters also play a role in the transition, such as stratification of ambient fluids (Hattori et al. 2013a; Pera & Gebhart 1971), the angle of the inclined plate (Alzwayi et al. 2014; Lewandowski 1991; Pera & Gebhart 1973), lateral width (Kalendar & Oosthuizen 2011), edge extension (Goldstein & Lau 1983) and plume formation point (Morton & Taylor 1956; Kimura et al. 2002).

The pioneering experiment presented by Eckert & Soehnghen (1951) revealed that a small natural disturbance could induce the amplification of the thermal boundary layer and trigger the transition from the laminar to the turbulent regime. The early linear stability theory was well established by Plapp (1957) and Szewczyk (1962) and this theory has been successfully used in experiment (Knowles 1968; Dring & Gebhart 1968; Polymeropoulos & Gebhart 1967). Furthermore, direct stability analysis has proved to be an efficient way to describe the flow on vertical plates and has been widely applied in many studies under different heating conditions (Javam & Armfield 2006; Ke et al. 2019; Zhao et al. 2013, 2019). The results showed that an infinitesimal disturbance near the leading edge might cause convective instability in the thermal boundary layer and the Bicoherence analysis helped to further investigate the transition from linear to transitional regime (Zhao et al. 2014). Meanwhile, the observation of longitudinal vortices on inclined plate (Haaland & Sparrow 1973; Lloyd & Sparrow 1970; Shaukatullah & Gebhart 1978; Sparrow & Husar 1969) showed a different transitional stage compared with the Tollmien-Schlichting (T-S) wave on the vertical plate. Following the onset of primary instability from 2D laminar to 3D laminar, the flow becomes periodic and finally enters turbulence in the downstream of the thermal boundary (Jeschke & Beer 2001; Shaukatullah & Gebhart 1978). By increasing the inclination angle from the horizontal plate to the vertical plate, the thermal boundary layer is separated downstream of the leading edge (Alzwayi et al. 2014; Paul et al. 2005). The occurrence of vortex instability has been proven to be an absolute instability and a disturbance dissipates soon after being ejected into the thermal boundary layer (Paul et al. 2005). The instability in the plume was further investigated. A 5% noise was applied near the heat source by Plourde et al. (2008), and the plume soon transited from a laminar regime to a turbulent regime. Except for the disturbance in constant heating condition, stratification also plays a significant role in transitional flow. Javam & Armfield (2006) revealed that a flow with a stratified ambient would exhibit bifurcation at a Rayleigh number when the homogeneous ambient flow does not exhibit bifurcation. Tao et al. (2004) and Tao & Busse (2009) indicated that the flow appears to have different transition routes in the thermal boundary layer near the inclined wall if stratified heating conditions and stratified ambient temperature were applied. Singh & Bajaj (2018) and Singh (2022) used Floquet analysis to investigate the instability response of inclined layer convection under time-dependent heating/cooling with temperature modulation. Harmonic response in longitudinal mode, harmonic, subharmonic and bicritical responses in transverse modes were found under different temperature modulations. Subsequently, the flow behaviour caused by Rayleigh–Taylor and Kelvin–Helmholtz instabilities was further analysed in Hattori et al. (2013a, 2013b, 2013c). Their results revealed that the forced convection in the background (i.e. due to the applied noise) reduces the effect of the Rayleigh–Taylor instability, and the bifurcation mode is almost inversed from a multi periods to single periodic mode. The occurrence of inverse bifurcation was also observed in a dynamic system, in which the increase of governing parameter leads the system to transit from a complex dynamical state to a simple state (Venkatesan & Lakshmanan 1998).

Furthermore, researchers have focused on the succession of bifurcation routes where linear stability analysis is difficult to achieve. Pallares & Grau (1999) presented different flow pattern transitions in the Rayleigh–Bénard laminar regime by flow topology for both numerical simulations and experiments. Continuous research (Puigjaner et al. 2004; Torres et al. 2013) based on the Galerkin spectral method produced a comprehensive bifurcation diagram under different Prandtl numbers and inclination angles. One of the usual transition routes is the Ruelle–Takens–Newhouse route (Ruelle & Takens 1971; Newhouse et al. 1978) with a quasi-periodic bifurcation based on dynamic system theory. In this route, the flow undergoes different bifurcation types with increasing control parameters such as Ra. Figure 1 illustrates predicted flow structures based on our recent computational fluid dynamics (CFD) modelling (Zhai et al. 2021). The primary bifurcation is a pitchfork bifurcation with the flow having a symmetry breaking, which means the stability changes when Ra is larger than critical value. Additionally, a flow transition into a periodic fluctuation state is found with a fundamental frequency $f_1$ that dominates the periodic flow, indicating the occurrence of a Hopf bifurcation. Further increasing Ra yields subharmonic frequencies (e.g. $f_1/2$) called period-doubling bifurcation. The appearance of more than one fundamental frequency (e.g. $f_1$ and $f_2$) in the power spectrum reveals that the flow undergoes a quasi-periodic bifurcation. With the occurrence of more fundamental frequencies, the flow becomes more complex and finally enters chaos. This bifurcation route has been widely observed in different systems (Bhowmick et al. 2019; Qiao et al. 2018a, 2018b, 2020; Zhai et al. 2021). In addition, subcritical transition (Vadasz 2000), period-doubling route (Feigenbaum 1980) and intermittency transition (Pomeau & Manneville 1980) also initiate other routes from steady to chaotic states.

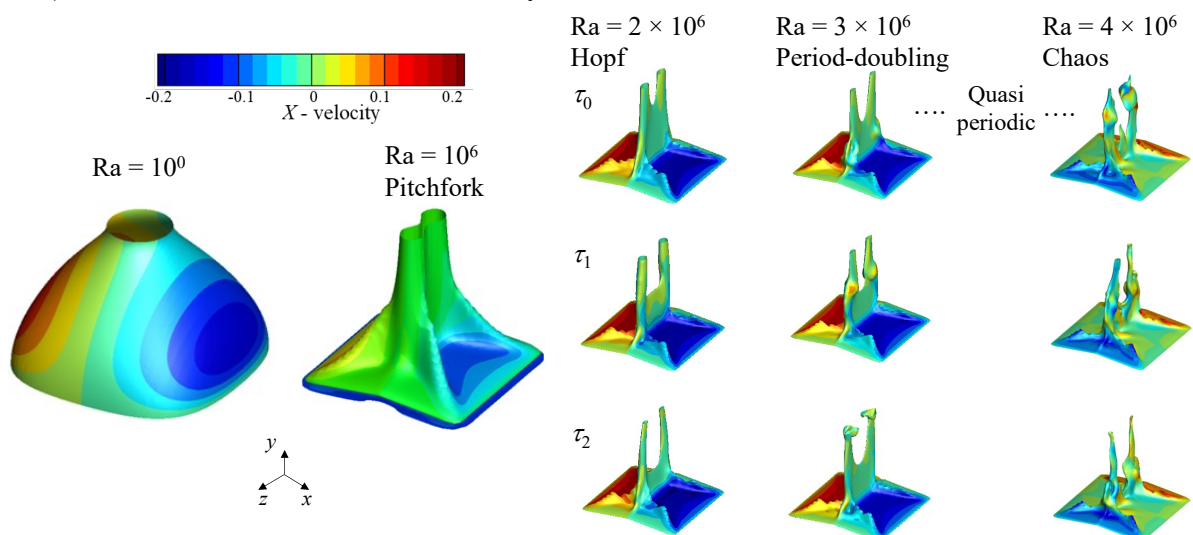

**FIG. 1**. Flow structure after different types of bifurcation that occur as Ra is increased over a section-triangular roof without thermal stratification in the surrounding environment (Zhai et al. 2021).

Our previous numerical study (Zhai et al. 2021) revealed a clear Ruelle–Takens–Newhouse transition route for an ideal fluid flow driven by a heated triangular roof (and undisturbed by external perturbations or 'background noise'). Unfortunately, such transition has not been observed experimentally. In this study, experimental measurements of flow transition thresholds and experimental characterisation of instability of the flow reveal potential effects of background noise and thermal stratification of the surrounding environment, which have not been considered numerically. Furthermore, as described in previous studies, some specific phenomena may exist during the transitional stage, e.g. occurrences of different transition routes (Javam & Armfield 2006), stability in advance (Plourde et al. 2008) and inverse bifurcation mode (Hattori et al. 2013c). Experimental data generally proves valuable for validating CFD models or improve theoretical frameworks. Therefore, in this study, we focus on the quantitative and qualitative descriptions of the thermal boundary layer structure and dynamics on the section-triangular roof. The instability occurrence is discussed and the bifurcation route with a critical Rayleigh number is analysed.

This paper starts presenting in Section 2 the experimental model, apparatus, and measurement procedures with important controlling parameters. Section 3 describes the development of the flow with increasing Rayleigh number. In subsection 3.1, bifurcation types in the unsteady state from laminar state to chaotic state are presented. In subsection 3.2, the flow mechanism in the chaotic state is discussed. In subsection 3.3, a discussion of a special transition route caused by ambient stratification is presented. The conclusions obtained from this experiment are presented in Section 4.

## 2. Experimental details

### 2.1. *Experimental model*

A section-triangular roof with slope 0.1 (A = $h/l$), and Prandtl number 0.71 (air) was considered. Schematics of the experimental models are illustrated in figure 2. To cover the wide range in Rayleigh number defined in subsection 2.3, two experimental models with different sizes (small and large) were constructed for this experiment. Figure 2(a) shows a small copper block (in orange colour) with a length of 60 mm length, width of 60 mm, and height of 23 mm. It was supported by an insulated stand (in yellow colour) made of thermoplastic polyester (polylactide). As illustrated in figure 2(b), a groove with a length of 60 mm, width of 60 mm and depth of 20 mm was located at the centre of the stand to place the copper block. Therefore, the part of the copper block excluding the groove can be considered as the section-triangular roof model. Thin insulation materials were smoothly attached to the sidewalls of the copper block to reduce heat exchange between the sidewall and the ambient. The copper block was designed to ensure that the Biot number was less than 0.1, such that uniform heating is applied to the surface of the roof during the experiment.

An active temperature control technique was applied to the small experimental model, where a Peltier module was placed underneath the copper block to heat and regulate the temperature of the roof during the experiment. The Peltier module is a device that functions based on the Peltier effect, which is the reverse phenomenon to the Seebeck effect (physical principle for thermocouples). Peltier modules enable heat flux control through metallic junctions whose materials have different Seebeck coefficients. The heat flux direction and intensity are controlled with the current direction and magnitude, respectively. This capability allows the Peltier module to reach the set temperature more accurately and rapidly compared to an ohmic heater. The heat flux of the Peltier module is controlled using a proportional–integral–derivative (PID) control system. By monitoring the temperature signal from a thermistor placed inside a drilled hole 2 mm below the surface, as shown in figure 2(b), the PID system can apply high power to the Peltier module to rapidly raise the temperature of the copper block. Real-time temperature control (heating or cooling) of the heated block is achieved through this system, effectively reducing overshooting when approaching the set temperature and minimising the impact of an overshoot on the development of the flow. A fin was used to extract heat from the Peltier module, allowing it to operate efficiently.

The large experimental model was a large aluminium block whose temperature was regulated by a water bath. As shown in figure 2(c), the aluminium block was 240 mm in length, 240 mm in width and 22 mm in height. A water tank was connected to a water bath with brass valves and hoses. To minimise the heat loss, the water tank was surrounded by foam insulation (not shown in the figure 2). The inner circulation in the water bath maintained the temperature of the experimental model. Similar to the small experimental model, a groove with a depth of 10 mm was also used to place the aluminium block, and the Biot number was small enough ($<$ 0.1) to guarantee uniform heating. Two V-shape fins were placed under the block as shown in figure 2(e). Holes were drilled on the fins, allowing the water to divert at different locations, thereby enhancing the heat exchange efficiency in the water tank. As illustrated in figure 2(d), two thermistors were placed inside the aluminium block to monitor the temperature and ensure its uniformity at different locations.

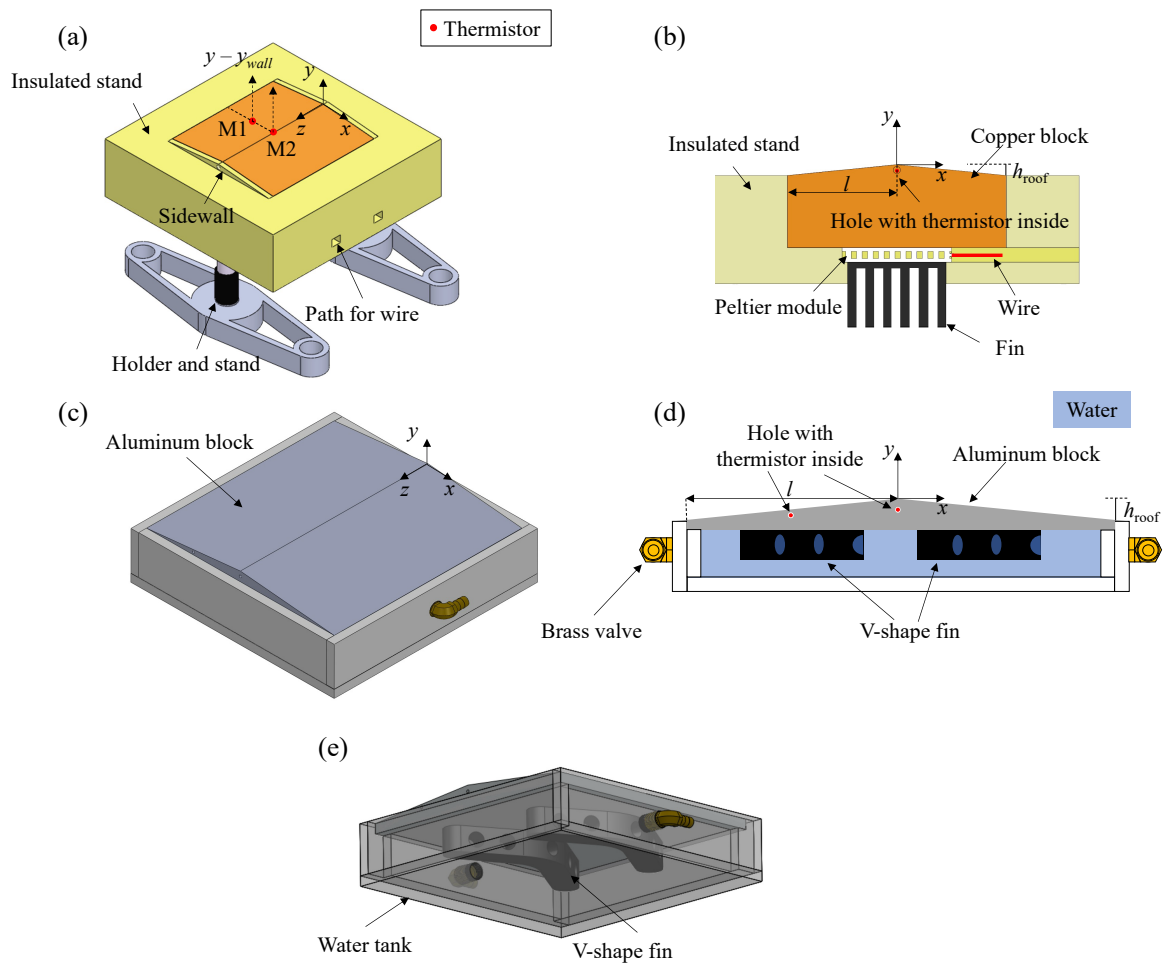

**FIG. 2**. Experiment models and heating systems. Schematics of (a) isometric view of the small experimental model, (b) front view of the small model, (c) isometric view of the large model, (d) cross-section view of the large model, and (e) bottom isometric view of the large model.

Calibrated thermistors were used to measure the temperature at different positions, as illustrated in figure 2(a). The response time of the thermistor is constant at ca. 0.1 s in air. The accuracy is 0.01 K, and the precision of measurement is 0.001 K. To achieve a rapid temperature measurement response with a minimum disturbance to the flow, the thermistor thermal mass was minimised, and its size limited to 1.2 mm × 1.6 mm. To obtain a non-invasive spatial-averaged temperature measurement (with the optical technique described in subsection 2.3), interferograms were taken with time intervals between each experiment of more than 1200 s to ensure that the flow is in a steady or quasi-steady state. Two thermistors, M1 and M2, were used to monitor the temperature development inside the thermal boundary layer and plume stem. M1 was located at $(-0.5l, l)$, and M2 was located at $(0, l)$ on the $x$–$z$ plane, as shown in figure 2(a). In different experimental tasks, these thermistors were mounted at an extension arm on a profile rail guide driven by a stepper motor, which precisely shifted the thermistor up and down to the desired height over the surface of the roof $(y - y_{wall})$, ranging from 2 mm to 10 mm $(y - y_{wall} = 2 \rightarrow 10$ mm$)$. The deviation was ca. 0.005 mm. To avoid downstream influences, these mounted thermistors were slightly staggered across the plate. Additional thermistors were placed in the block, water bath and container to measure the temperature in this experiment, e.g. three thermistors were placed inside the acrylic box shown in figure 3(b), one at the bottom and two at the top corners to measure the temperature in the environment surrounding the section-triangular roof.

## 2.2. *Experimental apparatus*

The flow visualisation experiment was performed based on a previously developed temporal phase-shifting interferometry (PSI) technique (Torres et al. 2012, 2020). This is because PSI can obtain high-resolution measurements of air flows adjacent to the heated surface; that is, particle image velocimetry (PIV) and shadowgraph in the authors' previous studies (Wang et al. 2018; Torres et al. 2020) were not adopted owing to the limitation of accurate and high-resolution temperature measurements of air flows on the heated surface. Additionally, thermistors were chosen because they can capture well the time dependent characterise of air flows on the heated surface.

As shown in figure 3(a), the laser beam was emitted from a 632 nm wavelength He-Ne laser source. The beam intensity was reduced after an ND filter, and the beam polarisation state was set with a linear polariser to 45° with respect to the horizontal plane. The spatial filter reduced the extra fringes and guaranteed that the beam smoothly formed a Gaussian distribution. The two concave and convex lenses behind it helped to obtain an expanded and collimated laser beam. The polarising beam splitter cube PBS-1 separated the beam into two beams (test and reference beams). After placing the experiment model in the test section shown in figure 3(b) within an acrylic box (to improve flow isolation from surrounding disturbances), the test beam affected by refractive index variations (on top of the heated roof) was merged with the undisturbed reference beam after passing through PBS-2. Then, a quarter-wave plate and a rotating polariser (driven by the stepper motor) produced circular and subsequent linear polarisation states, respectively. The vertical-plane temperature profile averaged along the optical path was obtained from the phase-shifted data as described by Torres et al. (2012). As shown in figure 3(b), a 500 mm cubic acrylic box covering the experimental model was used to help prevent undesired disturbances from the ambient air. A similar container was used by Saxena et al. (2020) and was shown to work well.

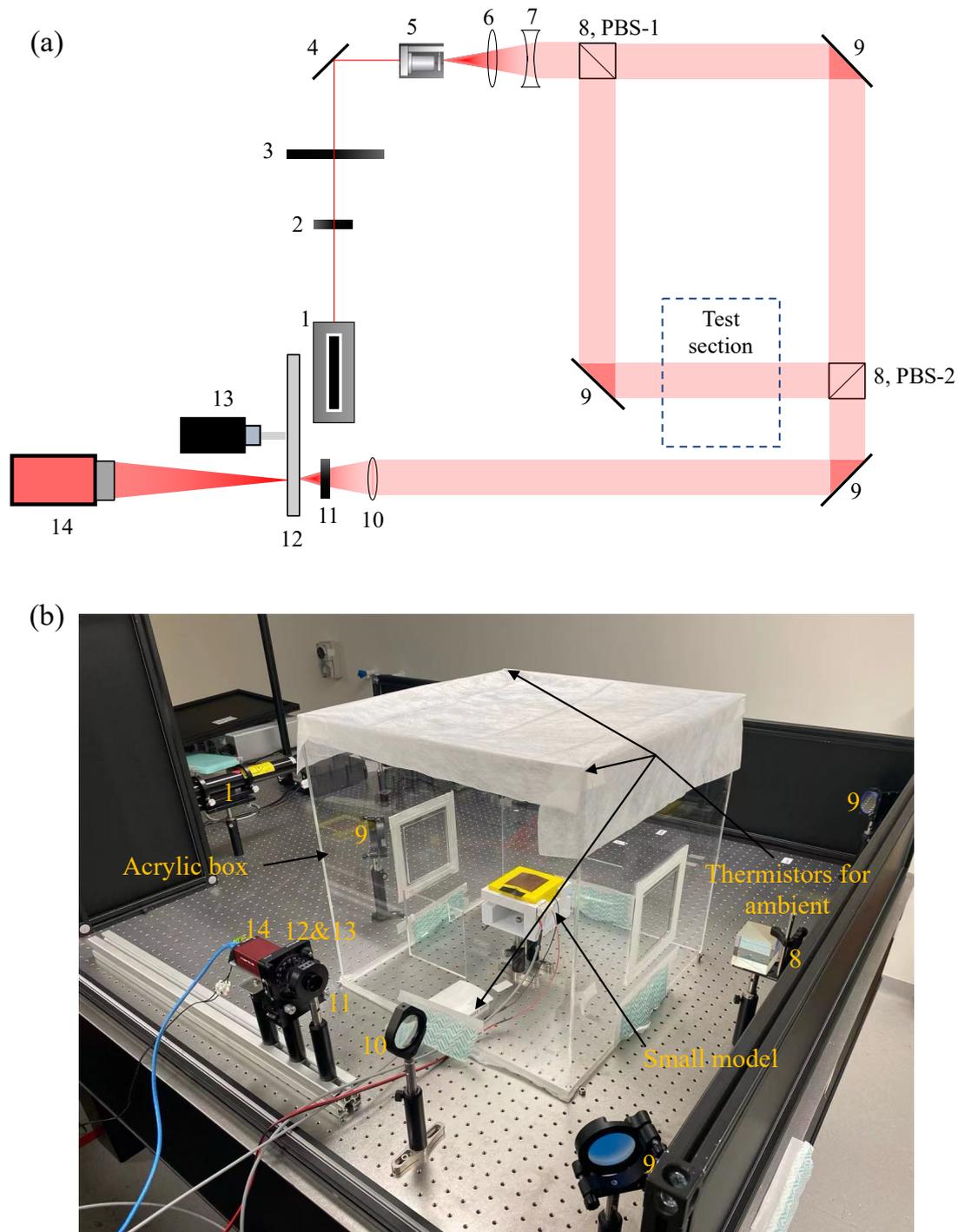

**FIG. 3**. Experimental setup. (a) Schematic of the phase-shifting interferometer with (1) 632 nm He-Ne laser, (2) ND filter, (3) linear polariser, (4) small reflection mirror, (5) spatial filter including objective lens and pinhole, (6) lens for beam expansion, (7) lens for beam collimation, (8) polarising beam splitter cube (PBS-1, PBS-2), (9) large reflection mirror, (10) focus lens, (11) quarter-wave plate, (12) rotating polariser, (13) stepper motor, and (14) CMOS camera. (b) Photo of experimental setup on the optical table.

### 2.3. *Experiment procedure*

A schematic of the laser beam passing through the small experiment model is illustrated in figure 4. The flow structures along both the $x$–$y$ and $y$–$z$ planes were visualised separately in different experiments with the same controlling parameters (also described in subsection 2.3). The beam area is approximately 20 mm × 20 mm. To achieve the highest quality of unwrapped data, some edge areas were cut off during post-processing; the size of the temperature field is indicated by the axes in each figure. The three areas shown in figure 4 were selected to visualise the thermal boundary layer and plume structure. Three thermistors (black dots in figures 4d, e and f) were placed slightly over the top boundary of the selected field of view or "Area" to measure the temperature used as boundary condition in our unwrapped phase-shifted data. To avoid spatial deviation, the thermistor would be placed at three typical positions (front, centre and back) along the beam direction and an average temperature would be obtained as a result. Area I, starting from half of the slope roof, focuses on the development of a thermal boundary layer on the inclined heated surface. Area II emphasises the structure of the plume stem. The boundary layer at the centre of the top surface was visualised using Area III. In contrast to the results in the $x$–$y$ plane, where the flow was averaged at the same height, the spatially averaged contour along the $y$–$z$ plane in Area III was acquired from the flow at different heights. In this case, it is better to describe the flow motion qualitatively. It is also worth noting that the positions of Areas I, II, and III in figure 4 are illustrated based on the size of the small experimental model.

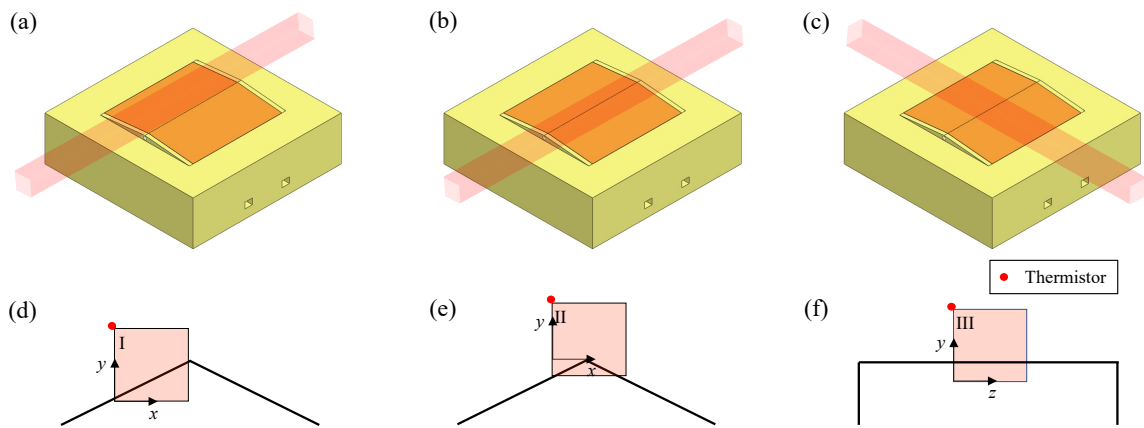

**FIG. 4**. Beam path in PSI and visualised areas. Schematic of the beam path on (a) the left side of the experimental model along $z$ direction, (b) the centre of the experimental model along $z$ direction and (c) the centre of the experimental model along $x$ direction. (d) – (f) Views along the beam direction corresponding to (a) – (c).

In this experiment, the raw data captured from PSI was further processed by using a three-bucket phase-shifting equation (Torres et al. 2012). Consecutive interferograms taken by the camera were transformed into phase-shifted data, as those shown in figures 5(a) and (b) which were taken from heating and no heating conditions, respectively. After that, an image processing procedure using two phase-shifted data was employed by Torres et al. (2020). The phase-shifted data in the non-isothermal (figure 5a) and isothermal (figure 5b; background) cases were unwrapped to produce the phase maps in figures 5(c) and (d), respectively. Then, the isothermal unwrapped data in figure 5(c) was subtracted from the background unwrapped data in figure 5(d) to obtain the phase map in figure 5(e). This method was used to determine the brightest pixel as the maximum temperature (heating temperature, $T_{heat}$) and the darkest pixel as the minimum temperature ($T_{min}$) in the experiment, and the actual temperature field with the $x$–$y$ axis is plotted in figure 5(f). The minimum temperature ($T_{min}$) can be measured directly with a temperature sensor (e.g. thermistor) when it is outside the thermal boundary layer, as illustrated in figures 4(d), (e) and (f). Therefore, the temperature resolution could be defined as

$$\text{Resolution} = \frac{T_{\text{heat}} - T_{\text{min}}}{N * 256}, \tag{3}$$

where $N$ represents the number of fringes (e.g. $N = 3.2$ in figure 5a) between these two points, noting that $N$ is not an integer number, i.e. it accounts for the number of fringes plus the remainder phase difference (Torres et al. 2020). Our custom-written code automatically calculated the specific size of $N$. '256' represents an 8-bit brightness representation range in decimals for each fringe. The selected coordinate system in figure 5 was calibrated by placing a ruler in the field of view before the experiment and using the lower left corner (illustrated in figures 4 d–f) as the coordinate origin. This calibration allowed for the determination of the specific dimensions of the sloping roof area at the bottom of the field of view and the overall height of the field, facilitating subsequent drawing and analysis.

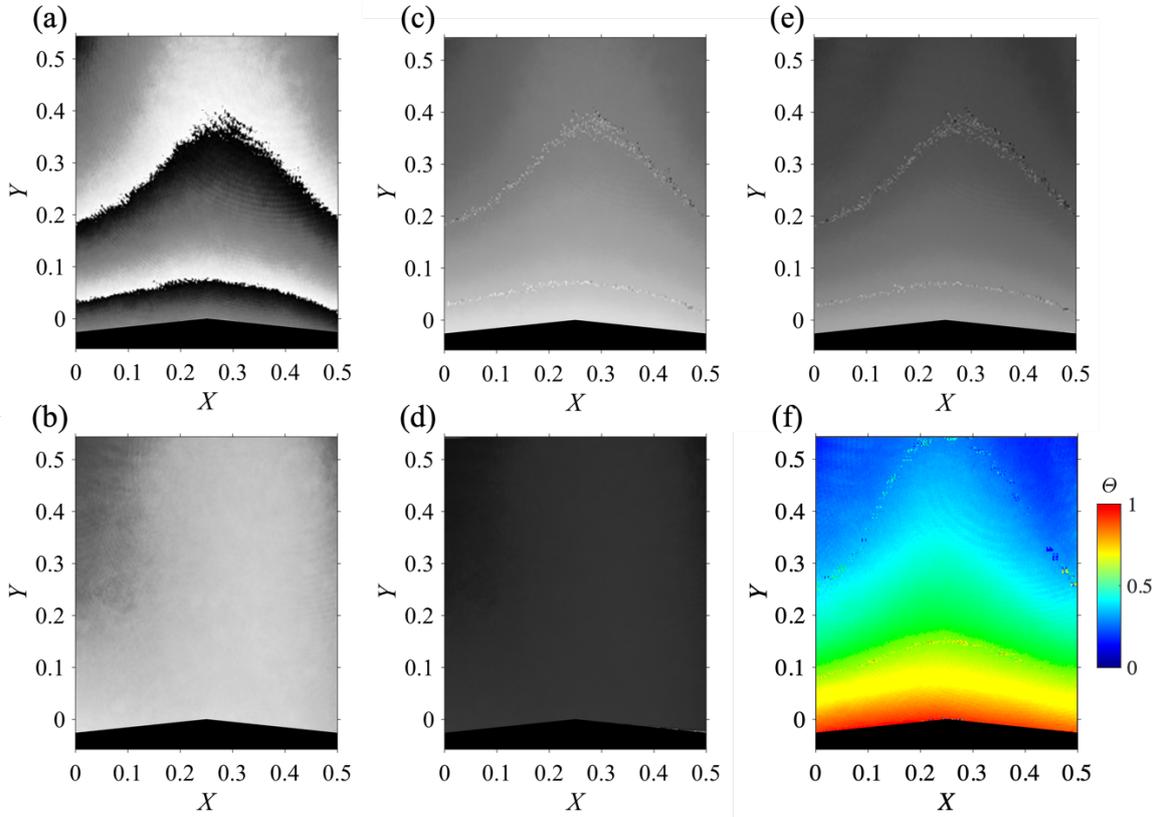

**FIG. 5**. Data processing for the image at Ra = $6 \times 10^4$. (a), (b) Wrapped phase-shifted data for non-isothermal flow and an isothermal background, respectively. (c), (d) Corresponding unwrapped phase map for (a) and (b). (e) Phase map after subtracting (d) from (c). (f) Measured temperature field. In the case, the experimental temperature difference is 22.6 K (also see case 15 in table 1).

Three nondimensional parameters were used as governing parameters in this experiment: the Rayleigh number (Ra), Prandtl number (Pr), and slope (A). Significant changes in flow dynamics and heat transfer may occur for the Rayleigh number ranging from $10^3$ to $4 \times 10^6$. In this study, $\Delta T$ in equation (1) is the temperature difference between the heated copper block and ambient air ($T_{\text{heat}}$ - $T_{\text{amb}}$). The Prandtl number and slope were fixed at 0.71 (air) and 0.1, respectively. The slope is defined as

$$A = h_{roof}/l, \tag{4}$$

where $h_{roof}$ is the height from the surface of the stand to the top of roof, and $l$ is the half width of the roof, as illustrated in figure 2. The entire experiment was roughly divided into two groups: $\Delta T$ in the range from 0.94 K to 38 K in experiment using the small copper block and $\Delta T$ in the range from 0.3 K to 24 K in experiment using the large aluminium block. The correlations between the characteristic length $l$, $\Delta T$ and Rayleigh number are presented in table 1. Except for the Rayleigh numbers listed below, experimental cases with critical Rayleigh numbers within the range were also performed in this study. A preliminary test showed that for those Rayleigh numbers larger than $4 \times 10^6$ in this experiment, the flow has entered a fully developed chaos, which is out of our scope in this study. Therefore, larger Rayleigh numbers were not further considered.

Time series of physical quantities at a specific point, such as temperature and velocity, are a common way to confirm the occurrence of bifurcation after the onset of single period bifurcation (Xu et al. 2008; Bhowmick et al. 2019; Li et al. 2020). Therefore, high-precision thermistors were adopted to measure these oscillations of air flows after the occurrence of single period bifurcation. In this study, temperature series in the fully developed state measured by thermistor M1 at $(x, z) = (-0.5l, l)$ and $y - y_{wall}$ ranging from 1 mm to 10 mm are selected to represent the near-field flow adjacent to the roof. Likewise, signals of thermistor M2 at $(x, z) = (0, l)$ and $y - y_{wall}$ ranging from 1 mm to 10 mm are selected to represent the behaviour of the downstream plume. To further investigate the flow behaviour, the power spectral density obtained by performing a fast Fourier transform of the temperature series was evaluated. It is worth noting that in the following sections, length scales $X$, $Y$, $Z$ are normalised by $l$, time scales in all figures are normalised by $l^2/\kappa Ra^{1/2}$, as shown in table 1. Therefore, the normalised time $\tau$ may differ 2–4 times with each other for a fixed duration (e.g. first 200 s in the fully developed state), which leads to an increased number of peaks and troughs in the graphs, potentially affecting readability. To prevent an excessive number of peaks and troughs and facilitate comparison and conversion for the readers, each figure is shown in an appropriate range of $\tau$. The temperature $\Theta$ is calculated as $(T - T_{amb})/\Delta T$, the height of thermistor $H$ is calculated by $(y - y_{wall})/l$, and the frequency $f$ in the power spectrum is dimensionless and normalised by $\kappa Ra^{1/2}/l^2$ based on a previous scaling analysis work (Zhai et al. 2018).

The fractal dimension $d_c$ is also calculated to better understand the transition state of the flow in quantity, which is defined as (Grassberger & Procaccia 1983):

$$d_c = \lim_{s \to 0} \frac{\log(c_p(s))}{\log(s)}, \tag{5}$$

$$C_p(s) = \lim_{N \to \infty} \frac{1}{N^2} \sum_{\substack{i,j=1 \\ i \neq j}}^{N} h(s - |\overrightarrow{X_i} - \overrightarrow{X_j}|), \tag{6}$$

where $h$ is the Heaviside function, $s$ is the maximum distance between $X_i$ and $X_j$, and $X_i$ and $X_j$ are the values of data points. In this study, $X_i$ and $X_j$ represented points in a three-dimensional phase space. The phase space was constructed based on time-delay embedding method which uses temperature signals from a single thermistor in the fully developed state by Grassberger & Procaccia (1983). For temperature signals obtained at different sampling times ($T_1$, $T_2$, $T_3$, …, $T_n$), every set of three consecutive data points was selected as one point $X_i$ in the phase space (e.g. $(T_1, T_2, T_3)$, $(T_2, T_3, T_4)$, …, $(T_{n-2}, T_{n-1}, T_n)$). It has been demonstrated that $d_c$ is close to 1 when single period bifurcation occurs. In a periodic transition state, including period-doubling bifurcation and quasi-periodic bifurcation, $d_c$ is between 1 and 2. When $d_c > 2$, it is believed that the flow enters chaos. However, it is still ambiguous to divide period-doubling bifurcation and quasi-periodic bifurcation by relying solely on the fractal dimension. In this case, the criterion of the transition state in this study depends on both the power spectrum and fractal dimension.

**Table 1**. Relationship between dimensional experimental parameters and the Rayleigh number, Ra. The characteristic lengths, some temperature differences and corresponding Ra in the two experimental models (figure 2) are listed. The factor $l^2/\kappa Ra^{1/2}$ is used to normalise the timescale.

| | Model roof size | Characteristic length, $l$ (mm) | Temperature difference, $\Delta T$ (K) | Ra | $l^2/\kappa Ra^{1/2}$ |
|---|---|---|---|---|---|
| 1 | 60 mm × 60 mm × 20 mm | 30 | 0.94 | $2.5 \times 10^3$ | 0.83 |
| 2 | | | 1.88 | $5 \times 10^3$ | 0.59 |
| 3 | | | 2.26 | $6 \times 10^3$ | 0.54 |
| 4 | | | 2.63 | $7 \times 10^3$ | 0.50 |
| 5 | | | 3.4 | $9 \times 10^3$ | 0.44 |
| 6 | | | 4.1 | $1.1 \times 10^4$ | 0.40 |
| 7 | | | 5.6 | $1.5 \times 10^4$ | 0.34 |
| 8 | | | 6.4 | $1.7 \times 10^3$ | 0.32 |
| 9 | | | 7.6 | $2 \times 10^4$ | 0.30 |
| 10 | | | 10.2 | $2.7 \times 10^3$ | 0.25 |
| 11 | | | 11.3 | $3 \times 10^3$ | 0.24 |
| 12 | | | 12.8 | $3.4 \times 10^4$ | 0.23 |
| 13 | | | 15.2 | $4 \times 10^4$ | 0.21 |
| 14 | | | 18.8 | $5 \times 10^4$ | 0.19 |
| 15 | | | 22.6 | $6 \times 10^4$ | 0.17 |
| 16 | 240 mm × 240 mm × 22 mm | 120 | 18 | $3 \times 10^6$ | 0.39 |
| 17 | | | 24 | $4 \times 10^6$ | 0.34 |

## 3. Results and discussion

### 3.1. *Flow bifurcation between unsteady state regimes*

First, consider the case at a relatively small Rayleigh number (Ra = $2.5 \times 10^3$). As shown in figure 6(a), the temperature of the near-field flow over roof is stationary accompanied by an intrinsic noise where the fluctuation range is less than the accuracy of the thermistor; therefore, the primary solution is quasi-steady. Second, when the Rayleigh number is increased to $5 \times 10^3$ as shown in figure 6(b), the flow started to oscillate within a very narrow temperature range where a single period was identified. The corresponding power spectrum analysis in figure 6(c) shows that the dominant frequency of this mono-periodic oscillation is $f_1 = 0.011$. The oscillating amplitude of the temperature time series significantly increased when increasing the Rayleigh number from $5 \times 10^3$ to $6 \times 10^3$, i.e. as shown in figures 6(b) and (d), respectively. Note that a subharmonic frequency of $f_1/2$ appeared in the power spectrum in figure 6(e). Although some additional frequencies appeared which might be caused by background disturbances and experiment accuracy (the increase of heating temperature is small), this still indicates that a period-doubling bifurcation occurred between $5 \times 10^3$ to $6 \times 10^3$. The flow developed with multiple periods at Ra = $9 \times 10^3$, as shown in figures 6(f) and (g). Except for the characteristic frequency $f_1$ which slightly decreases to 0.0094 and its harmonic frequencies, an incommensurable frequency $f_2$ which equals to 0.0037 was observed. The corresponding fractal dimension was found to be 1.07, which means that a quasi-periodic bifurcation occurred between $6 \times 10^3$ and $9 \times 10^3$. When the Rayleigh number increased to $1.1 \times 10^4$ in figure 6(h), it resulted in temperature oscillations with amplitudes much greater than those observed in the oscillatory flow in figure 6(d). The flow fluctuated in a series of irregular and intermittent bursts. As shown in figure 6(i), it is difficult to distinguish the primary peak frequency. The occurrence of multiple main frequencies and fractal dimension results ($d_c = 1.46$) reveals that the flow undergoes further quasi-periodic bifurcation and is very close to chaos.

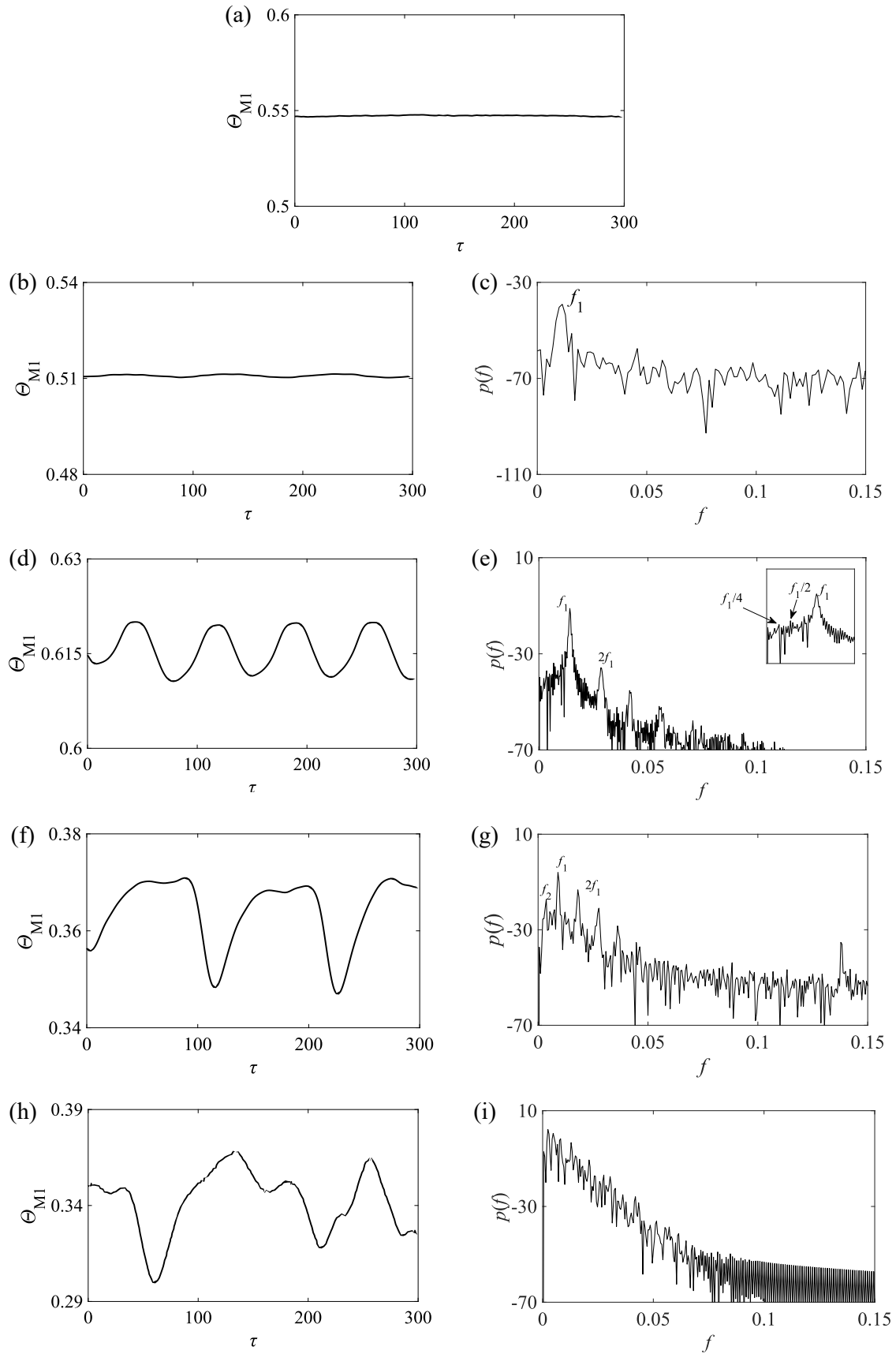

**FIG. 6**. Temperature series of monitor point M1 at the $H = 0.03$ for (a) Ra $= 2.5 \times 10^3$, (b) Ra $= 5 \times 10^3$, (d) Ra $= 6 \times 10^3$, (f) Ra $= 9 \times 10^3$ and (h) Ra $= 1.1 \times 10^4$ with their corresponding power spectrum in (c), (e), (g) and (i), respectively.

In the range of the Rayleigh number described above, the spatial-averaged temperature contour from the PSI technique was also observed. A previous study (Zhai et al. 2021) has revealed that the flow along a three-dimensional inclined plate still has a quasi-two-dimensional structure in the conduction dominance regime and early convection dominance regime. Therefore, the spatially averaged flow contour along the spanwise direction ($Z$, $X$–$Y$ plane) was first investigated in Area II (refer to figure 4 for Area definitions). When the Rayleigh number is smaller than $10^4$, the weak flow shown in figure 7 was dominated by conduction, as there was no distinct thermal boundary layer in the slope roof. Instead, only a dome-like structure was present. This measurement agrees well with previously reported numerical simulation results (Zhai et al. 2021). It is worth noting that the thermal boundary layers processed from different experiments cannot perfectly fit with each other, especially at a relatively low Rayleigh number. This is because the temperature difference in the contour is small and the temperature gradient in the contour is very sensitive. However, the flow in every experiment for Ra = $10^4$ still appeared to be a uniformly distributed temperature contour. When the flow is of a higher Rayleigh number, the areas in different experiments gradually fit well.

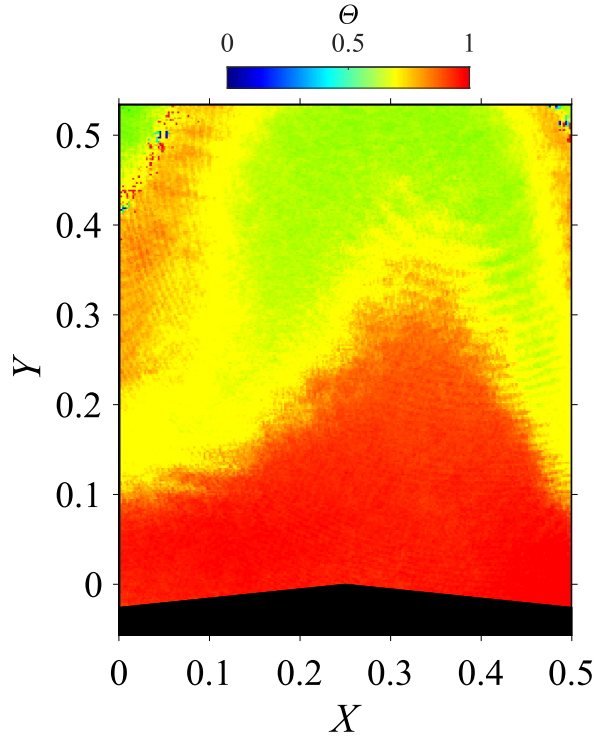

**FIG. 7**. Temperature contour spatially averaged along $Z$ on Area II for Ra = $10^4$.

As described above, the temperature fluctuated periodically for Ra = $6 \times 10^3$ near the heated surface ($H = 0.03$), as shown in figure 6(d). The corresponding power spectrum contains only a fundamental frequency $f_1$, together with its harmonic and subharmonic frequencies, as shown in figure 6(e). In figure 8, the time series of near-field flow inside the thermal boundary layer on Ra = $6 \times 10^3$ at $H = 0.15$ are also shown. It can be observed in figure 8(a) that the crest of the temperature series at $H = 0.15$ started to collapse with twice the amplitude of the temperature series at $H = 0.03$. Similar with figures 6(d) and (e), the peak frequency $f_1$ and its harmonic frequencies in figure 8(b) clearly show the power spectral density of the temperature series. Note that some additional frequencies were measured by thermistor, but these could be attributed to background disturbances and measurement precision.

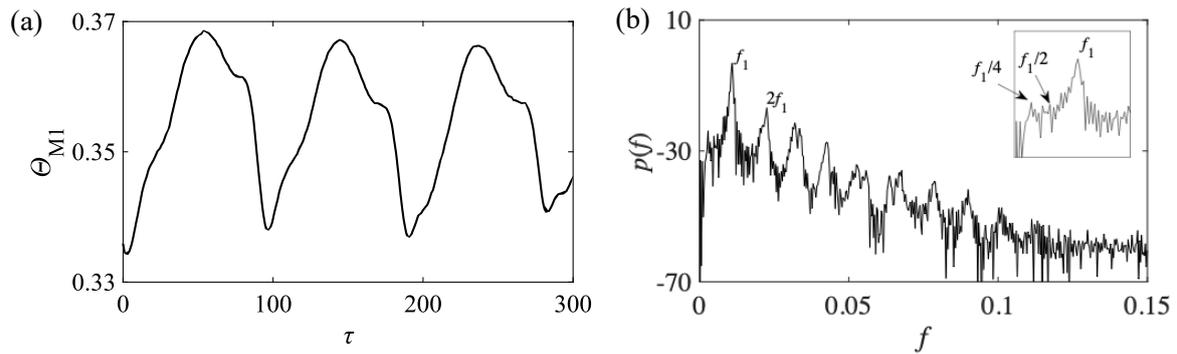

**FIG. 8**. Temperature series at monitor point M1 for Ra = 6 × 10³ at the height of (a) $H$ = 0.15. The power spectrum corresponding to (a) is plotted in (b).

In the meantime, the near-field flow also developed along the inclined roof. Comparing the measurement results inside plume stem at $H$ = 0.03 in figure 9(a) and at $H$ = 0.15 in figure 9(c) which are both located at the top of the roof for Ra = 6 × 10³, the increasing amplitude of the oscillatory flow indicates a gradual loss of stability from the upstream to the downstream flow, resembling a typical convective instability. The power spectral densities in figures 9(b) and (d) have $f_1$ and its corresponding harmonic and subharmonic frequencies. In this case, as is discussed in figure 6 and figure 8, the flow still gradually transits from period-doubling to quasi-periodic mode in the plume stem.

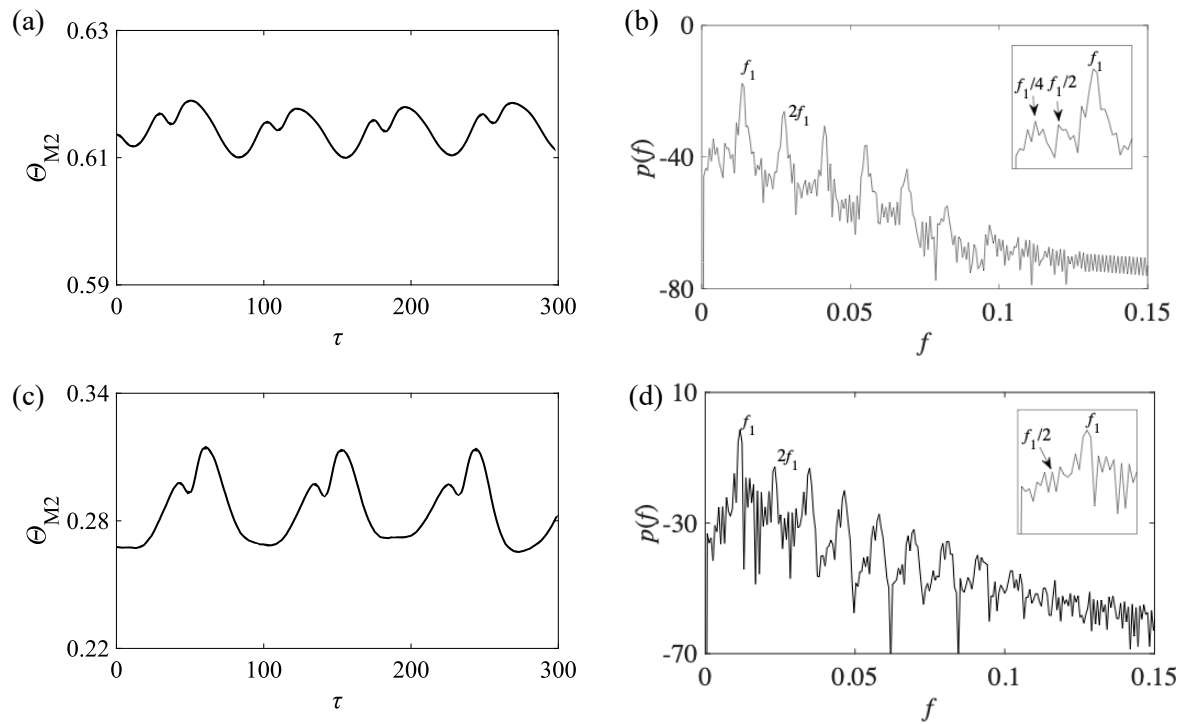

**FIG. 9**. Temperature history at monitor point M2 for Ra = 6 × 10³ at height (a) $H$ = 0.03 and (c) $H$ = 0.15. The power spectra corresponding to (a) and (b) are plotted in (c) and (d), respectively.

Further investigation of the downstream far from the near-field flow is beyond the scope of this study. Instead, the temperature development inside the plume stem was investigated further with an increase in the Rayleigh number. As shown in figure 10(a), an oscillatory flow was observed. The power spectral density illustrated in figure 10(b) indicates that only the fundamental frequency $f_1$ distinctly exists. This indicates that these anomalous frequencies observed in figure 8 and 9 might have been caused by experiment inaccuracy. As shown in figure 10(a,c), the amplitude increased from 0.015 to 0.025 when the height $H$ increased from $H = 0.03$ to 0.15. Additionally, the corresponding fractal dimension value is 0.977. Considering the power spectral density and fractal dimension, it is likely that the flow is still under period-doubling bifurcation but very close to quasi-periodic bifurcation because of the additional frequency occurrence.

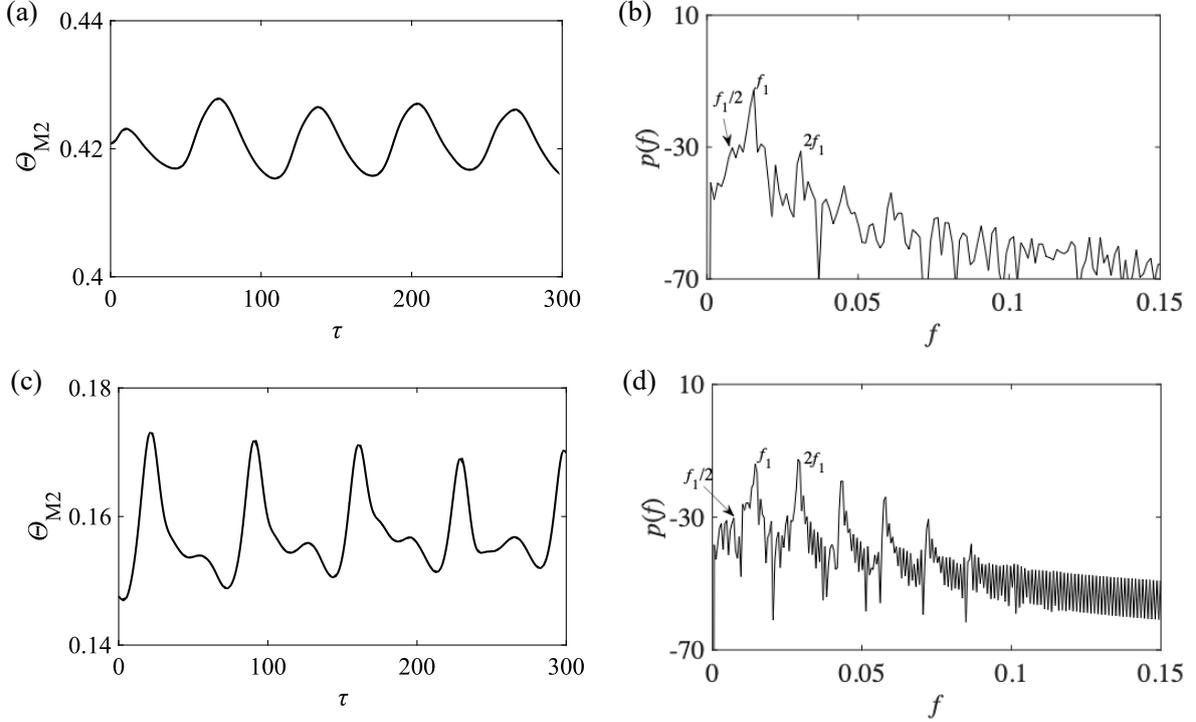

**FIG. 10**. Temperature history monitor point M2 for Ra = $7 \times 10^3$ at height (a) $H = 0.03$ and (b) $H = 0.15$. The corresponding power spectrum for (a) and (b) is plotted in (c) and (d).

As shown in figure 11(a), for Ra = $1.5 \times 10^4$, the oscillation of the near-field flow was minimised, and the primary peak frequency ($f_1 = 0.002$), which is much lower than that in figure 6(e), was obtained based on figure 11(b). This indicates that an inverse single period bifurcation occurred between $1.1 \times 10^4$ and $1.5 \times 10^4$. The fluctuation grew again for Ra = $1.7 \times 10^4$, as shown in figure 11(c). The characteristic frequencies obtained in the power spectrum (figure 11c) are $f_1 = 0.012$ and $f_2 = 0.004$ (which is two times larger than the peak frequency at Ra = $1.5 \times 10^4$). This indicates that the peak frequency gradually changed from 0.002 (or its harmonic frequency 0.004) to a higher frequency of 0.012. When the Rayleigh number reached $2.3 \times 10^4$, as shown in figure 11(e), the flow was greatly amplified. Based on the power spectrum in figure 6(f), there is no distinct peak frequency, and the flow is in the intermittent chaotic state. However, the result presented in figure 11(g) shows a further development of quasi-periodic mode with its harmonic frequency and additional frequencies. After that, the flow entered chaos at Ra = $3.4 \times 10^4$, supported by the typical continuous broadband power spectrum in figure 11(j) and the fractal dimension result ($d_c = 2.03$). Some periodic fluctuations can still be observed in this figure. This is because the early chaotic state is still far from fully developed turbulence and still contains some periodic flows.

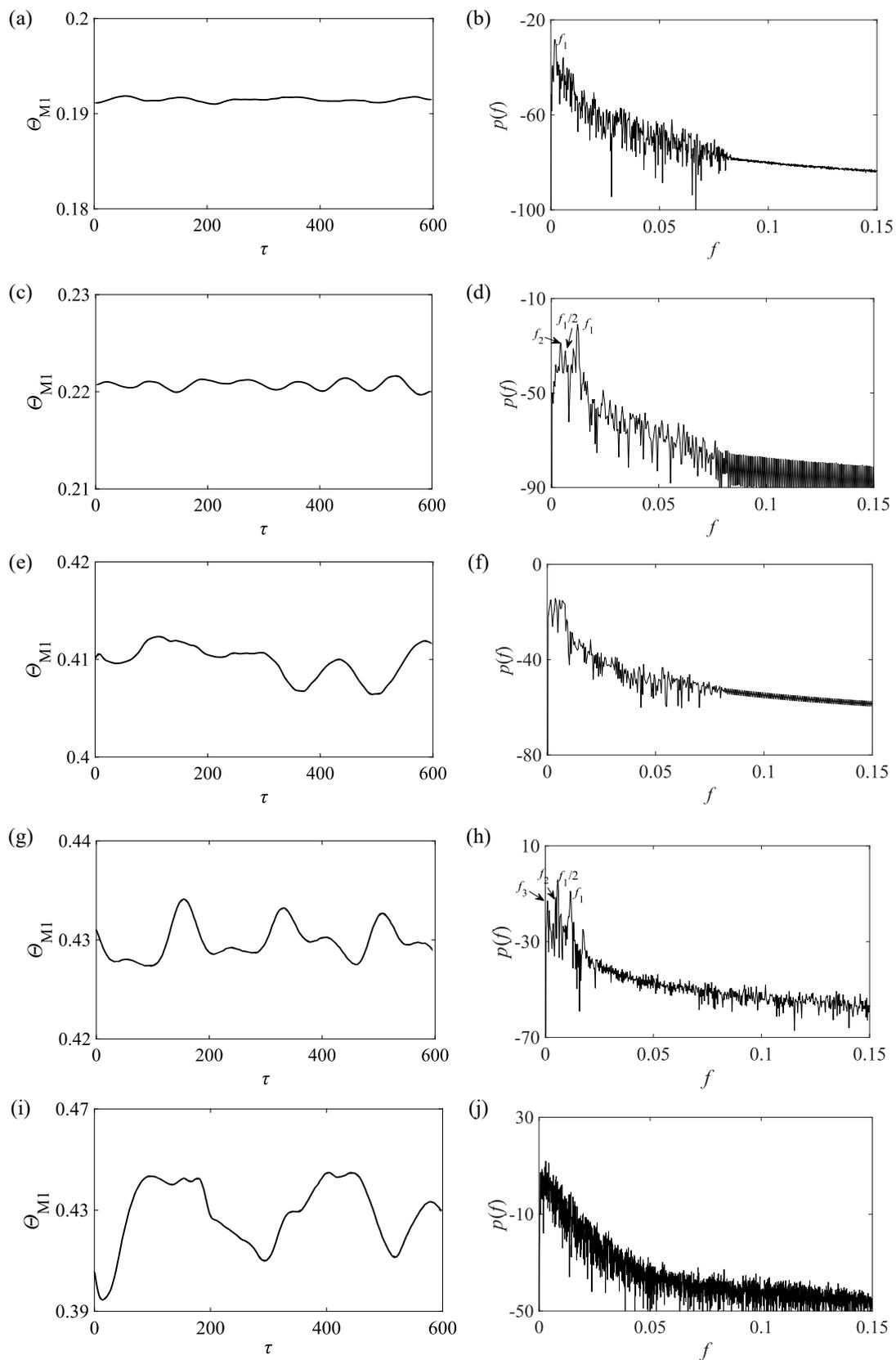

**FIG. 11.** Temperature series of monitor point M1 at the height of $H = 0.03$ in (a) Ra = 1.5 × 10⁴, (c) Ra = 1.7 × 10⁴, (e) Ra = 2.3 × 10⁴, (g) Ra = 2.7 × 10⁴ and (i) Ra = 3.4 × 10⁴ with their corresponding power spectrum in (b), (d), (f), (h) and (j), respectively.

With an increase in the Rayleigh number from $1.5 \times 10^4$ to $3.4 \times 10^4$, the temperature at the monitor point gradually increased, as shown in figure 11. Near the heated surface, characterised by relatively hot air (depicted in red and yellow), a discernible temperature gradient was developing from $Y = 0$ to $0.35$ along the line $X = 0.25$. This indicates the gradual formation of a thermal boundary layer with an increase of convection. However, for the relatively cold air (as is depicted in blue in figure), contours over $Y = 0.42$ along the line $X = 0.25$ in figure 12 remained uniformly distributed. The thickness of the flow was larger than the beam area, resembling the primary structure in figure 1. This indicates that when the Rayleigh number reached $3.4 \times 10^4$, the flow is still under conduction dominance without a clear plume stem, as shown in figure 12.

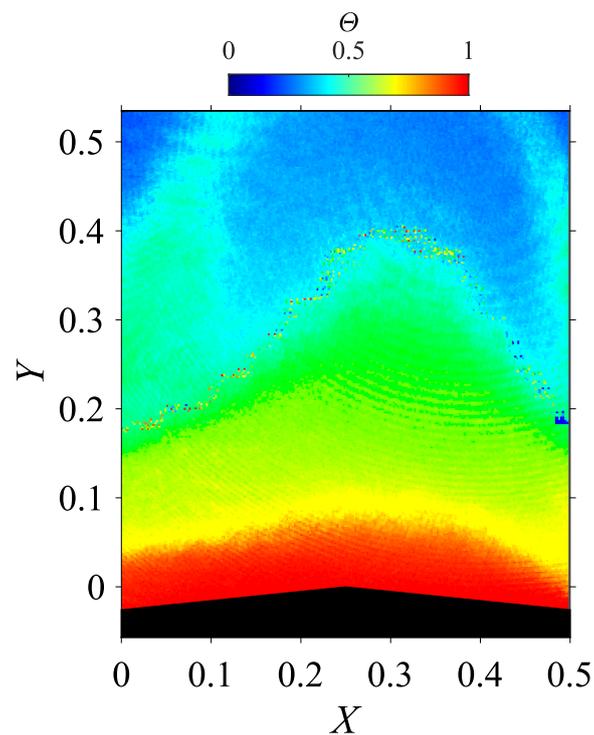

**FIG. 12**. Temperature contour spatially averaged along $Z$ on Area II for Ra = $3 \times 10^4$.

### 3.2. *Flow transition into chaos*

When the Rayleigh number reached $4 \times 10^4$, the fluctuation of the flow in figure 13(a) was slightly reduced compared with the result in figure 11(i). However, by further increasing the Rayleigh number to $5 \times 10^4$ (shown in figure 13c) or larger, the flow started to amplify again. Meantime, the corresponding power spectra in figures 13(b) and (d) indicate that the flow was in a fully developed chaotic state.

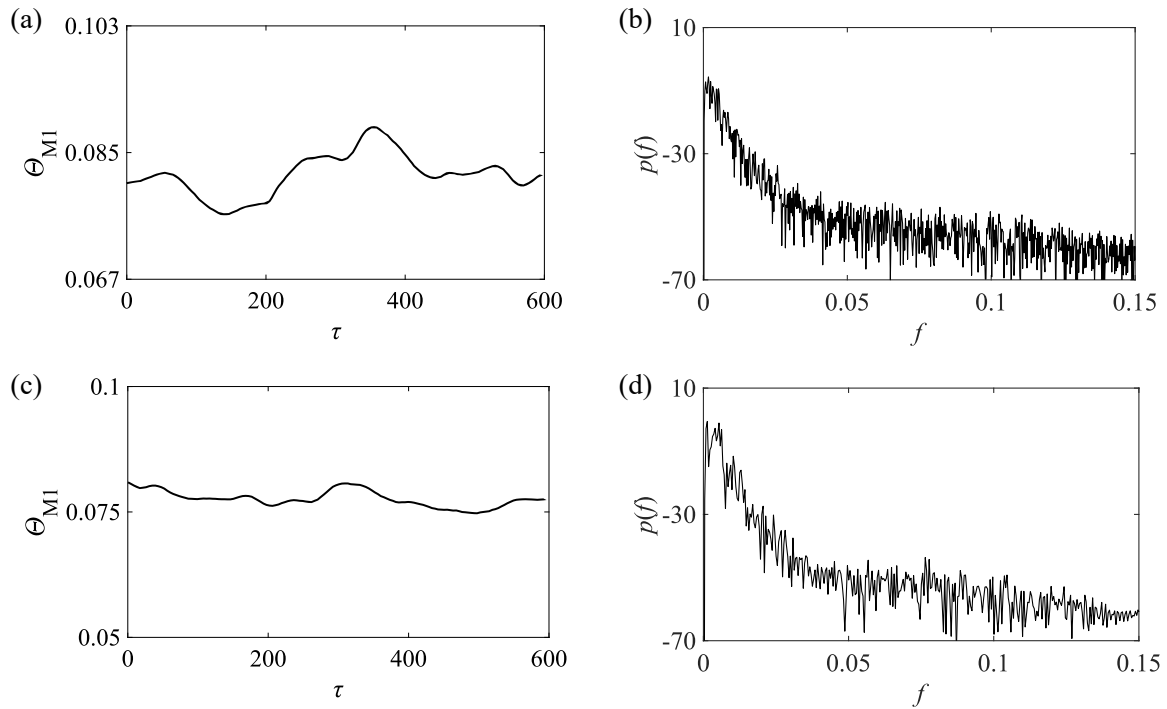

**FIG. 13**. Temperature series of monitor point M1 at the height of $H = 0.03$ for (a) Ra = 4 × $10^4$ and (c) Ra = 5 × $10^4$ with their corresponding power spectrum in (b), and (d), respectively.

After reaching the fully developed chaotic state, the temperature contour from the PSI technique at Ra = 6 × $10^4$ is analysed. As depicted in figure 14(a), the thermal boundary layer became distinct, and a plume stem formed around $Y = 0.45$. Compared with our previous numerical results (Zhai et al. 2021) shown in figure 14(b), the temperature profile in the visualised contour from the PSI technique has excellent agreement.

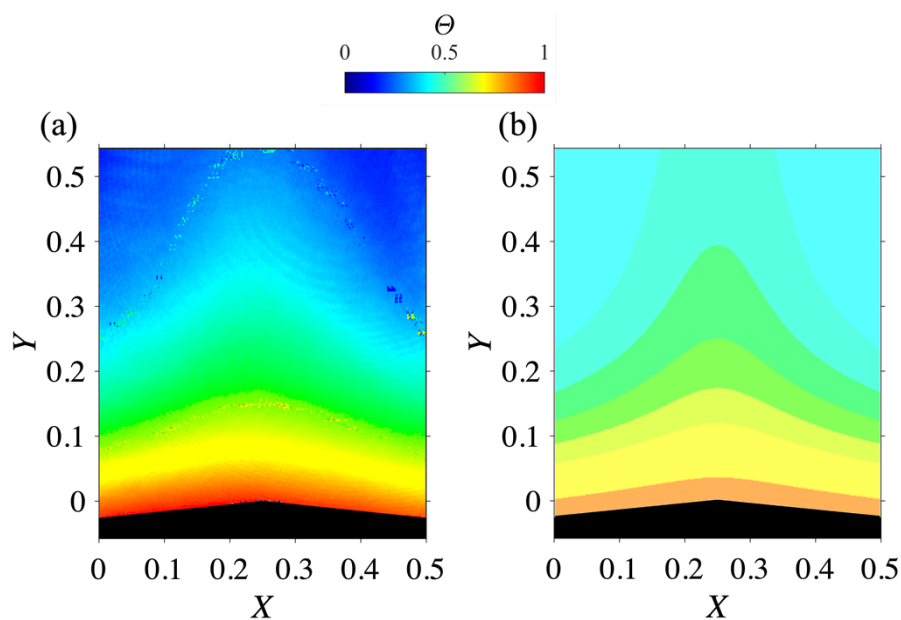

**FIG. 14**. Comparison between (a) PSI-based experimental results in this work and (b) the temperature contour spatially averaged along $Z$ on Area II for Ra = 6 × $10^4$ based

on our previously reported simulations (Zhai et al. 2021).

To further understand the flow structure, visualised contours in Area III are also analysed in both the unsteady state (mainly described in subsection 2.1) and chaotic state. As shown in figure 15(a), for a relatively low Rayleigh number of Ra = $10^4$, the temperature contour over the roof appeared as a dome-like structure and is symmetric along the line $Z = 0.2$. Considering figure 7 and figure 15(a), it is believed that the flow structure is centrosymmetric. However, the dome-like structure in Area III is smaller than that in Area II. This is because the laser beam was partly blocked by the roof, and the hot air near the surface could not be captured. When the Rayleigh number gradually increases to $3 \times 10^4$ in figure 15(b), the thermal boundary layer formed over the surface and tilted slightly from left to right. Further increasing to $6 \times 10^4$, as shown in figure 15(c), the thermal boundary layer became thinner, which means that the convection effect dominates heat transfer of the flow.

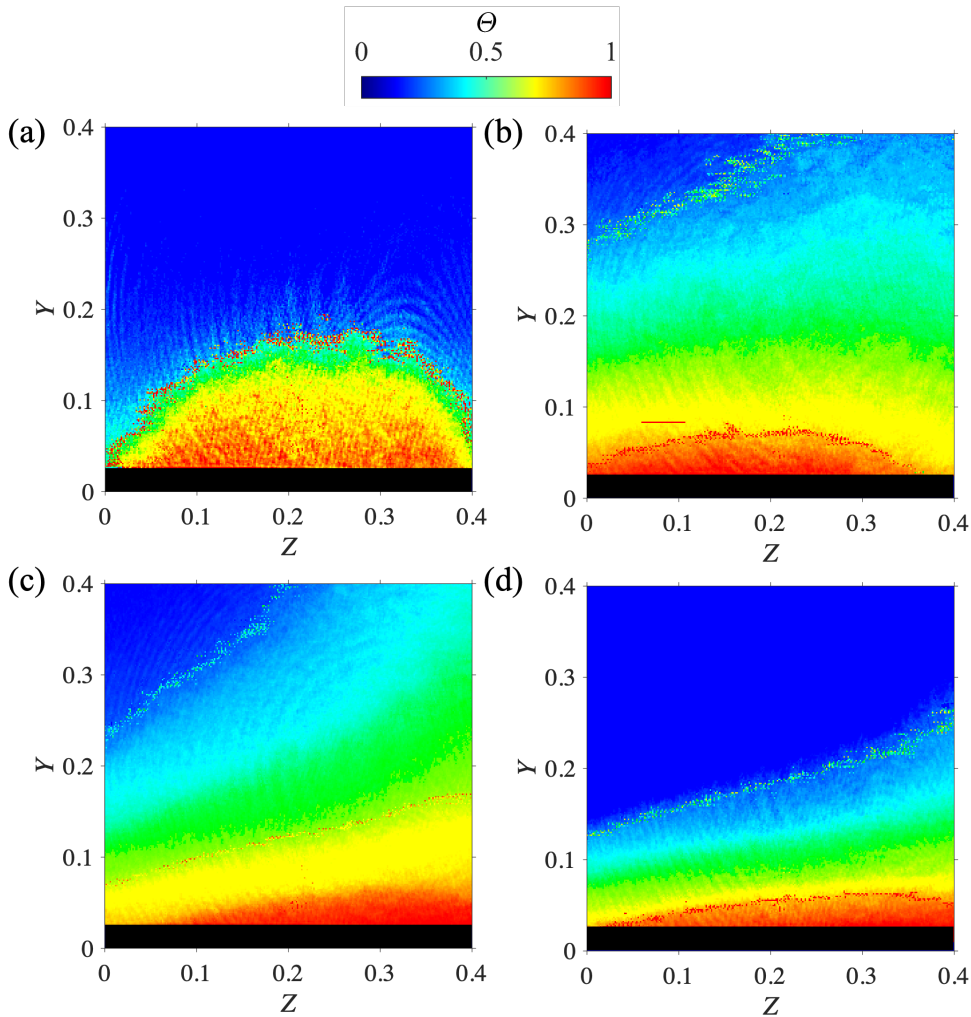

**FIG. 15**. Temperature contour spatially averaged along $X$ on Area III. (a) Ra = $10^4$, (b) Ra = $3 \times 10^4$, (c) Ra = $5 \times 10^4$, and (d) Ra = $6 \times 10^4$.

After entering the fully developed chaotic state, a travelling wave occurred in Area I, as shown in figure 16(a). As indicated in figures 16(a)–(d), the travelling wave passes through Area I within $\tau = 2.07$. Subsequently, a new travelling wave occurred at $\tau = 4.14$, as shown in figure 16(e). When the travelling wave reached Area II, figure 16(f) reveals that the plume stem was vertically upward at first and then tilted to the right, influenced by the travelling wave from upstream, as shown in figure 16(g). After the travelling wave passing through the top of the roof, the plume stem broke and disappeared, as shown in figures 16(h)–(j).

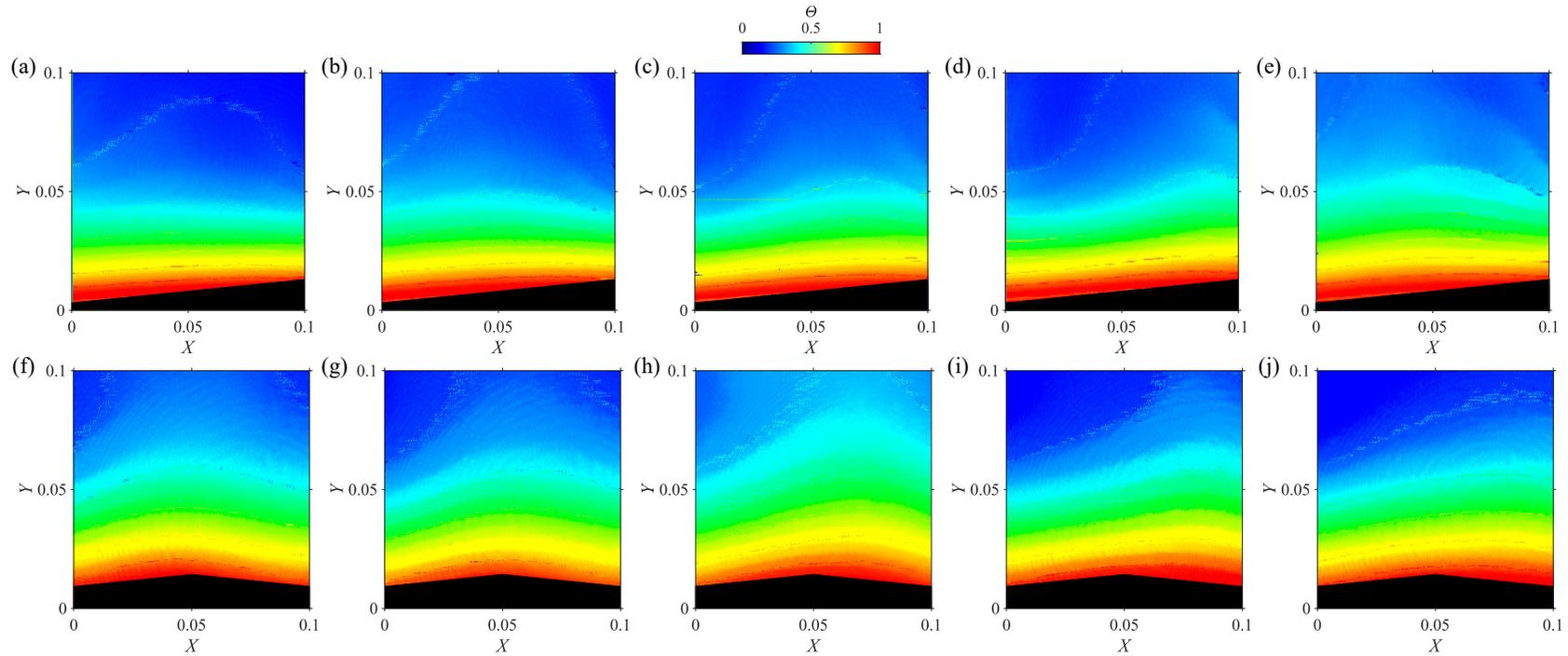

**FIG. 16**. Temperature contours spatially averaged along $z$ direction for Ra = $3 \times 10^6$ in Area I (top) and Area II (bottom) at (a), (f) $\tau = 0$, (b), (g) $\tau = 2.07$, (c), (h) $\tau = 4.14$, (d), (i) $\tau = 6.21$ and (e), (j) $\tau = 8.28$.

### 3.3. *Discussion: transition route with increasing Rayleigh number*

The fractal dimension and positions of some critical temperature series are plotted in figure 17. For the bifurcation route we discussed above, the fractal dimension value at Ra = 5 × 10³ (marked as (a) in figure 17) supports the conclusion that the single period bifurcation occurs. The fractal dimension value at Ra = 6 × 10³ (figure 17b) indicates that a period-doubling bifurcation might exist before the flow reached the quasi-periodic bifurcation at 9 × 10³ (figure 17c). When Ra is increased to Ra = 1.1 × 10⁴, a fully developed quasi-periodic bifurcation is observed at the fractal dimension $d_c = 1.464$. However, when Ra = 1.5 × 10⁴ (figure 17d), the nonlinearity is reduced and the fractal dimension value decreases. The flow reverses to the single period bifurcation again. After that, the flow becomes increasingly close to instability from Ra = 2.7 × 10⁴ (figure 17e) and finally enters chaos at Ra = 3.4 × 10⁴ (figure 17f). Although some fractal dimension values are still lower than 2, their corresponding power spectrum results reveal that the flow is in the chaotic state.

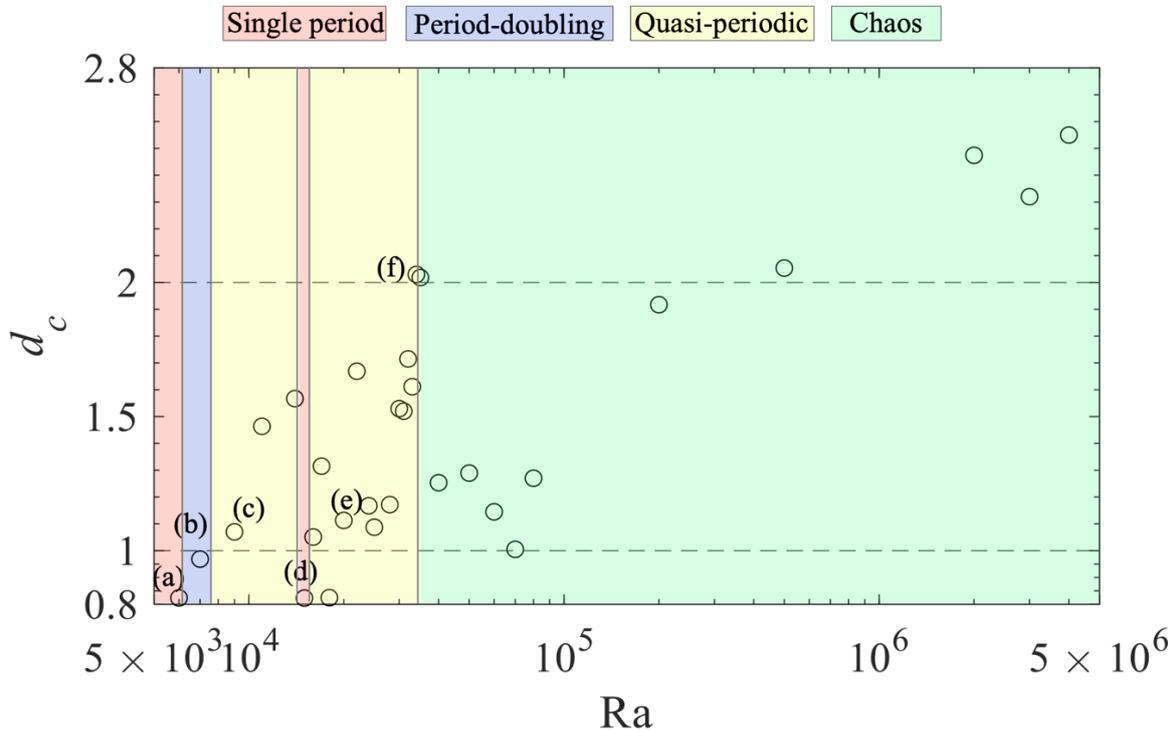

**FIG. 17**. Fractal dimension value (circle point) at location M1 for different Rayleigh numbers varying from 5 × 10³ to 4 × 10⁶. Typical phase spaces were observed at (a) Ra = 5 × 10³, (b) Ra = 6 × 10³, (c) Ra = 9 × 10³, (d) Ra = 1.5 × 10⁴, (e) Ra = 2.7 × 10⁴ and (f) Ra = 3.4 × 10⁴.

To further understand the transition route in this study, the primary frequencies $f_1$ for different Rayleigh numbers are shown in figure 18. It is observed that the primary frequency remained constant before the first occurrence of quasi-periodic bifurcation (red solid line) where the frequency slightly decreased to 0.009. Upon further increasing the Rayleigh number, the frequency continuously decreased from 0.009 to 0.002. Until now, the flow was still under quasi-periodic bifurcation. After that, the flow occurred inverse single period bifurcation shown in figure 17(d) and then transited to quasi-periodic bifurcation again within a very narrow range shown in figure 17(e) and the frequency turns back to 0.012. It is worth noting that the harmonic frequency of the primary frequency at Ra = 1.5 × 10⁴ can still be observed at Ra = 1.7 × 10⁴. Finally, the flow entered chaos at Ra = 3.4 × 10⁴, where some peak frequencies could still be observed, as indicated by the blue line.

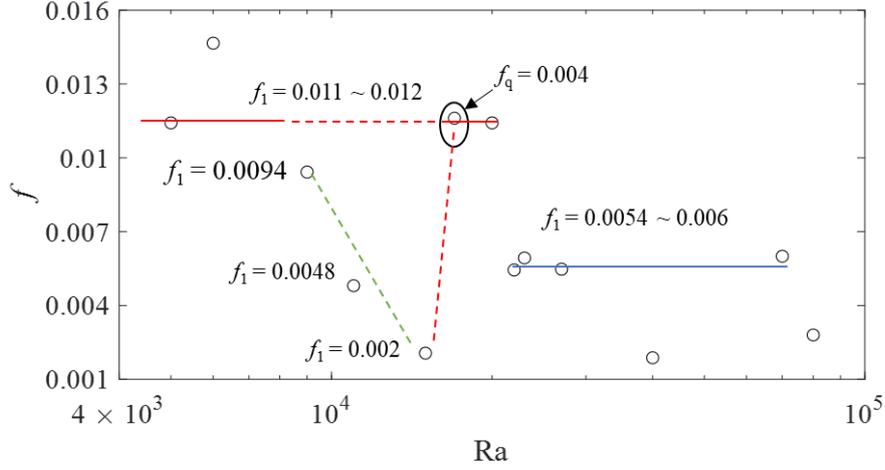

**FIG. 18**. Evolution of primary frequency when increasing the Rayleigh number from 5 × 10³ to 10⁵.

Compared with our previous numerical results (Zhai et al. 2021), natural convection over a roof with no stratified background undergoes periodic oscillation at Ra = 2 × 10⁶ with a fundamental frequency $f_1 = 0.21$. It is evident that the characteristic Rayleigh number in the numerical simulation is several orders of magnitude higher than that in the experiment. In this experiment, we observed that the bottom and top of the container have a temperature difference of approximately 0.3 K to 0.5 K temperature difference, which is the background stratification in the laboratory. In addition, the fundamental frequency in the numerical simulation lies in a relatively high-frequency band, whereas the frequency in the experiment lies in the lower-frequency band. As investigated by Javam & Armfield (2006), the stratified background induces shear instability and a thermal boundary layer near the vertical wall, and the flow first appears as a low-frequency bifurcation signal in the air. In this study, we found that the fundamental frequency decreases linearly with the increment in the Rayleigh number, as a relation of $f \sim$ Ra⁻¹, indicated by the green dashed line in figure 18. Considering the normalised relation $\kappa$Ra$^{1/2}/l^2$, the dimensional result is:

$$F/(\kappa/l^2) \sim \text{Ra}^{-1/2}, \qquad (4)$$

where $F$ denotes the dimensional frequency. This result corresponds to the internal wave mode in a side-open cavity described by Javam & Armfield (2006). In this case, it is concluded that the flow development before the occurrence of the inverse single period bifurcation can be identified as a low-frequency mode caused by an internal wave in the stratified background air inside the cavity of this study. Based on figure 18, we can also see that the high-frequency signal first occurs as the harmonic mode of the low-frequency signal. In addition, the PSI visualised contours reveal that the flow structure still appears symmetric, as shown in figure 14, where the flow has entered chaos. This indicates that the instability caused by the internal wave mode is weak. As shown in figure 16, a travelling wave was clearly observed by the PSI technique, indicating that the high-frequency mode from travelling waves finally caused the instability of the convective system, breaking its flow symmetry.

## 4. Conclusions

A complex bifurcation route was experimentally investigated using a previously developed temporal phase-shifting interferometer technique (PSI) and temperature measurement method. In this experiment, we focused on the flow development in a wide range of Rayleigh numbers from $10^3$ to $4 \times 10^6$ at a constant Prandtl number and slope of 0.71 and 0.1, respectively.

At first, the near-field flow remained steady under conduction dominance when Ra < $2.5 \times 10^3$ and the flow structure was axisymmetric in both the $X$–$Y$ and $Y$–$Z$ planes. By slightly increasing Ra to $5 \times 10^3$, the oscillatory flow showed the occurrence of single period bifurcation. Subsequently, a period-doubling bifurcation near Ra = $6 \times 10^3$ and a quasi-periodic bifurcation near $9 \times 10^3$ occurred continuously in a very narrow range. Between Ra = $1.5 \times 10^4$ and $1.7 \times 10^4$, a reverse transition from fully developed quasi-periodic bifurcation to single period bifurcation occurred, where the convection effect became increasingly strong. With the increase in Ra, the flow entered the quasi-periodic bifurcation at Ra = $2.7 \times 10^4$ and finally reached chaos when Ra = $3.4 \times 10^4$.

The spatially averaged counters were also quantitatively and qualitatively described. The flow structures developed from the quasi-steady state to the chaotic state were also visualised using the PSI technique. In the conduction dominant regime, there was no distinct thermal boundary layer. With an increase in the Rayleigh number, the thermal boundary layer grew together with the effect of convection. Although flow oscillation occurs in this regime, the structure still appears symmetric and finally becomes asymmetric after the flow enters a fully developed chaotic state.

Through the results obtained from phase-shifting interferometry (PSI) and thermistors, we have revealed the evolution of different flow modes in a thermally stratified environment, highlighting the low-frequency transition induced by internal waves and the high-frequency transition caused by traveling waves. We have characterised the bifurcation phenomena observed in the low-frequency transition and further discussed the development of flow on the variation of spatial position within the boundary layer and plume stem. In conclusion, we summarised the bifurcation phenomena occurring within the near-field flow as the Rayleigh number increases, as well as the underlying principles governing the generation of flow dynamics and the transformation processes of low-frequency and high-frequency waves with varying Rayleigh numbers.

Previous studies provided a separate focus on the flow dynamics under different slope (Pera & Gebhart 1971), Prandtl number (Hattori et al. 2013a), and thermal stratification gradients (Javam & Armfield 2006; Krizhevsky et al. 1996). The experiments reported in this study spark a compelling interest in further analysing flow transition under varying parameters that dictate flow behaviour. For example, it will be interesting to further investigate whether these findings extrapolate under coupled parameters, e.g. different Prandtl numbers, slope, thermal stratification gradients and lengths in spanwise direction.

## Acknowledgements

This work was supported financially by the National Natural Science Foundation of China (Grant No. 11972072). Additionally, simulations were undertaken with the assistance of resources and services from the National Computational Infrastructure, which is supported by the Australian Government; funding from the 2019 ANU Global Research Partnership Scheme for the project "Exchange on multiscale convective heat transfer" is also acknowledged.

## Declaration of Interests

The authors report no conflict of interest.